\documentclass[aip, sd, amsmath,amssymb,reprint]{revtex4-1}
\usepackage{graphicx}
\usepackage{dcolumn}
\usepackage{bm}
\usepackage{graphicx}
\usepackage{amsmath}
\usepackage{mathptm} 
\usepackage{hyperref}
\usepackage{xcolor}
\preprint{APS/123-QED}
\usepackage[normalem]{ulem}

\begin{document}
\preprint{APS/123-QED}
\title{Amplitude chimera and chimera death induced by external agents in two-layer networks}%
\author{Umesh Kumar Verma and G. Ambika}
\affiliation{Indian Institute of Science Education and Research(IISER) Tirupati, Tirupati, 517507, India}

\date{\today}

\begin{abstract}
We report the emergence of stable amplitude chimeras and chimera death in a two-layer network where one layer has an ensemble of identical nonlinear oscillators interacting directly through local coupling and indirectly through dynamic agents that form the second layer. The nonlocality in the interaction among the dynamic agents in the second layer induces different types of chimera related dynamical states in the first layer. The amplitude chimeras developed in them are found to be extremely stable, while chimera death states are prevalent for increased coupling strengths. The results presented are for a system of coupled Stuart-Landau oscillators and can in general represent systems with short-range interactions coupled to another set of systems with long range interactions. In this case, by tuning the range of interactions among the oscillators or the coupling strength between the two types of systems, we can control the nature of chimera states and the system can also be restored to homogeneous steady states. The dynamic agents interacting nonlocally with long-range interactions can be considered as a dynamic environment or medium interacting with the system. We indicate how the second layer can act as a reinforcement mechanism on the first layer under various possible interactions for desirable effects.
\end{abstract}

\maketitle

\begin{quotation}

Chimera states are emergent dynamical patterns in a network of coupled identical oscillators where coherent and incoherent domains coexist. There is growing evidence that the study of chimera states can help to understand the behavior of many real-world systems. Most of the studies on chimera states are in single networks. Recently interactions of chimera states across coupled layers in multilayer networks~\cite{Maksimenko, Majhi, Majhi1, Sawicki} are reported. Such studies on multilayer networks, deal with systems where each layer has the same type of dynamics at its nodes. In this work, we study the dynamics of a two layer network where first layer has an ensemble of identical nonlinear oscillators with local or short-range interactions, and the second has systems with a different  nodal dynamics and nonlocal or long-range interactions among them. We consider the second layer to be dynamic agents that can also function as a dynamic environment in interaction with the network of systems in the first layer. We show how the nonlocality in the interactions of the second layer can induce chimeras and control related dynamics in the first layer. We observe stable amplitude chimera (AC) for weak interlayer coupling, and as interlayer strength increases, we observe chimera death (CD) and other different types of steady-states such homogeneous steady-state (HSS), inhomogeneous steady-state (IHSS), 2-cluster steady-state (2-CSS), and multi-cluster steady-state (MCSS).

\end{quotation}

\section{Introduction}

The study of complex systems using the framework of complex networks has attracted a lot of attention in recent research in many areas ~\cite{watts}. The emergent behavior in such systems due to interaction among the dynamical units reveals a variety of interesting cooperative phenomena, such as synchronization~\cite{Pikovsky}, suppression of oscillations~\cite{Saxena,Koseska}, chimera~\cite{Kuramoto}, amplitude chimera~\cite{Koseska2014}, chimera death~\cite{Koseska2014, Banerjee}, etc. Among these, synchronization is the most widely studied one, and it broadly deals with the transition from incoherence to coherence among coupled dynamical systems. The suppression of oscillations observed in such systems is another emergent phenomenon, which can be classified into two, namely, amplitude death(AD)~\cite{Saxena} and oscillation death(OD)~\cite{Koseska}. In AD, coupled oscillators settle at a common stable steady-state, which is the fixed point of the uncoupled system, while OD, refers to the situation where the final state is a new coupling-dependent steady state(s). In this case, coupled oscillators may settle to different steady states [termed inhomogeneous steady states (IHSS)], or to a homogeneous steady-state (HSS).

The chimera state is an interesting spatiotemporal behavior where spatially coherent and incoherent behavior of oscillators coexist in a network of coupled identical oscillators. Kuramoto and Battogtokh first observed this peculiar spatiotemporal pattern in a network of phase oscillators with a simple symmetric nonlocal coupling scheme~\cite{Kuramoto}, and later this was mathematically established by Strogatz et al.~\cite{Abrams}. Subsequently chimeras were found in periodic oscillators~\cite{Ulonska}, chaotic oscillators~\cite{Gu}, chaotic maps~\cite{Omelchenko},  time-delay systems~\cite{Gopal} and  neuronal systems which exhibit bursting dynamics~\cite{Bera,Chouzouris}. Initially, chimera states were reported in nonlocally coupled systems, but later it was also found in globally coupled systems~\cite{Yeldesbay, Chandrasekar}, locally coupled systems~\cite{Laing, Hizanidis}, indirectly coupled systems~\cite{Chandrasekar2016, Gopal2018}, and modular networks~\cite{Hizanidis2016}. Besides numerical and theoretical studies, chimera patterns have also been demonstrated in laboratory experiments. In particular, chimera patterns were observed in an electro-optical system~\cite{Tinsley,Hagerstrom}, mechanical systems~\cite{Martens}, chemical oscillators~\cite{Nkomo},  electrochemical systems~\cite{Wickramasinghe, Wickramasinghe2014}, electronic circuits~\cite{Gambuzza, Rosin}, and optical combs~\cite{Viktorov}. Depending on the initial conditions and network topology, various types of chimera states are observed on networks, such as amplitude mediated chimera~\cite{Sethia}, amplitude chimera~\cite{Koseska2014}, chimera death~\cite{Koseska2014, Banerjee}, globally clustered chimera~\cite{Sheeba}, phase-flip chimera~\cite{Chandrasekar2016, Gopal2018},  imperfect chimera~\cite{Kapitaniak}, imperfect traveling chimera~\cite{Bera2016}, breathing chimera~\cite{Abrams} etc. 

In addition to its established wide prevalence, chimera states are found to play an important role in the various dynamical behaviors of many real-world systems. For example, in the case of aquatic animals like dolphins and migratory birds, unihemispheric slow-wave sleep is a phenomenon where only one hemisphere of the brain shows sleep activity. The sleeping part of the brain exhibits highly synchronized activity while awake part of the brain shows desynchronized activity~\cite{Rottenberg}. During epileptic seizures, a part of the brain remains highly synchronized, while the remaining part is desynchronized~\cite{Rothkegel}. Chimera states have also been linked to the various types of brain diseases such as Alzheimer's disease,  Parkinson's disease, schizophrenia, and brain tumors~\cite{Uhlhaas}. The interplay of synchrony and asynchrony as displayed by chimera states plays an important role in brain function and disease as reported in recent studies. Stationary moving chimeras are seen in network of FitzHugh-Nagumo neurons with empirical structural brain network topology and simulated modular fractal topology\cite{Chouzouris}. 

Most of the real world systems are not isolated but interact among themselves as well as with their environment or external systems. Such an environment can be modelled as a system  of coupled elements where all the elements communicate to each other through dynamical agents or signalling molecules, that can freely diffuse in the surrounding medium. Examples of such systems include genetic oscillators~\cite{Kuznetsov},  chemical oscillators~\cite{Toth}, and ensemble of cold atoms~\cite{Javaloyes}, etc. There are several studies that are focused on the various collective dynamics possible in oscillators that interact with each other through a dynamic environment~\cite{Katriel,Resmi,Resmi2012,Ghosh,Sharma2016,Sharma2016a,Verma,Verma2018,Verma2019,Verma2019b}. 

The study of multilayer networks is a recent topic of research that has relevance in understanding the dynamics of several complex systems, like multilayer structures in neural networks~\cite{Maksimenko, Majhi, Majhi1, Sawicki}. We present the framework of multilayer networks to study the interaction between two ensembles of systems, of which one layer has nonlinear oscillators with local or short-range interactions, and the other has systems of a different dynamics with nonlocal or long-range interactions. We take the second layer to be dynamic agents that can together function as a dynamic environment in interaction with the network of systems in the first layer. We study how the nonlocality in the interactions of the second layer can induce chimeras and control related dynamics in the first layer, when both layers are connected in a feed back loop. Thus our model is different from most of the recently studied models where both layers have identical dynamical systems~\cite{Maksimenko, Majhi, Majhi1, Sawicki}.

We report how the network of locally connected identical oscillators splits into coexisting coherent and incoherent domains due to the influence of the environment having nonlocal interactions. For weak interlayer coupling strength, we see stable amplitude chimera (AC) and as this coupling strength increases, chimera death (CD) and other different types of steady-states occur. Interestingly, emergent dynamics of the oscillators can be controlled by tuning the range of interactions in the environment layer and we report a variety states like stable amplitude chimera, chimera death, HSS, IHSS, 2-cluster steady state(2-CSS), multi cluster steady state(MCSS), 1-state chimera death, 2-state chimera death and travelling waves as possible emergent dynamical states. we note amplitude chimera state  is found to stabilize  through nonlocal repulsive coupling in the presence of attractive coupling in a system of oscillators on a regular network, even for random initial conditions~\cite{Sathiyadevi1,Sathiyadevi2}.  Our results are in two layer regular networks, each layer with different nodal dynamics and  the mechanism of creation of chimeras in the present study is thus due to feedback from another layer that is nonlocal in connectivity.

\begin{figure}
\includegraphics[width=0.45\textwidth]{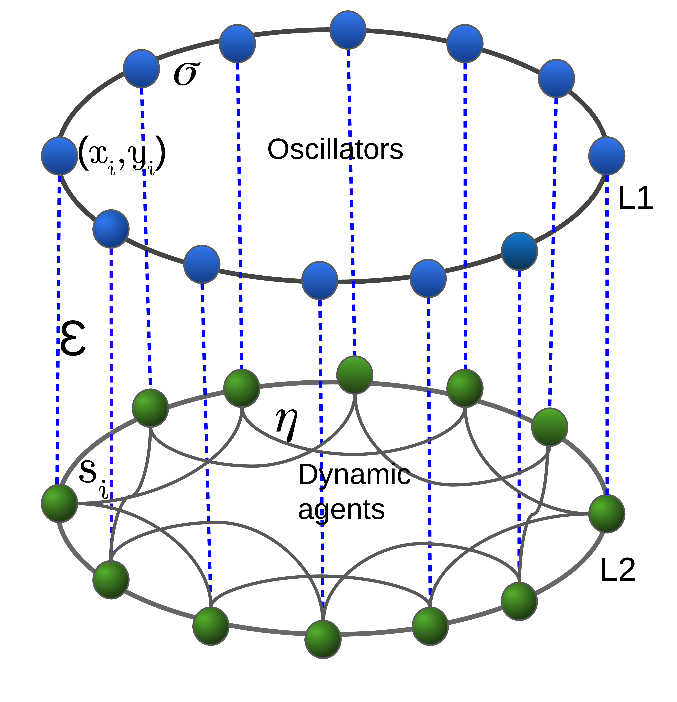}
\caption{Schematic diagram of the two layer network where nodes in upper layer L1 (blue) represent the dynamics of oscillators and that the lower layer L2(green) describe the dynamic agents. Each oscillator in L1 is connected to the corresponding dynamic agent in L2.}
\label{fig1}
\end{figure}

\section{Interacting two layer networks}
The multilayer network under study consists of two layers, as shown schematically in Fig.~\ref{fig1}. The top layer consists of an ensemble of $N$ Stuart-Landau oscillators (SL), with local intralayer diffusive coupling, called system layer, L1. They have interlayer feedback coupling with the dynamic agents in the second layer, called L2, with multiplex like i to i coupling. The dynamic agents are 1-d overdamped oscillators with intralayer diffusive couplings that can model the presence of an interacting environment or medium. Their dynamics is sustained due to feedback from L1 but can in turn influence the dynamics on L1 through the feedback coupling. The dynamics of the two-layer network thus modelled is given by
\begin{eqnarray}
\dot{x}_{i}&=&(1-x_{i}^2-y_{i}^2)x_{i}-\omega y_{i}+\frac{\sigma}{2P_1}\sum_{j=i-P_1}^{i+P_1}(x_{j}-x_{i})+\epsilon s_i \nonumber\\
\dot{y}_{i}&=&(1-x_{i}^2-y_{i}^2)y_{i}+\omega x_{i} \nonumber\\
\dot{s}_i &=& -\gamma s_i-\epsilon x_{i}+\frac{\eta}{2P_2}\sum_{j=i-P_2}^{i+P_2}(s_j-s_i)
\label{eq1}
\end{eqnarray}

\noindent where $i=1,2,\ldots,N$. $x_i$ and $y_i$ are the state variables of the $i^{th}$ Stuart-Landau(SL) oscillator. The individual SL oscillator exhibits limit cycle oscillations with natural frequency $\omega$. The $i^{th}$ oscillator interact with other oscillators directly and indirectly through dynamic agent $s_i$ on th other layer with feedback coupling of strength $\epsilon$. The dynamics of the dynamic agents $s_i$ is considered to be one-dimensional over-damped oscillators with damping coefficient $\gamma>0$. 
The interaction between the oscillators in first layer is controlled by $\sigma$ and $P_1$, whereas the interaction between dynamic agents in second layer is controlled by $\eta$ and $P_2$. $ P_1 \text{ and } P_2, \in \{1, N/2\}$, correspond to the number of nearest neighbors in each direction on each layer respectively. They thus represent the range of interaction with the coupling radius defined by $R=\frac{P}{N}$. For local coupling $P=1$, for global coupling  $P=\frac{N}{2}$ and for nonlocal coupling value of $P$ is in the range $1<P<N/2$.  By varying $ P_1 \text{ and } P_2 $, we can study the influence of nonlocality in coupling on the dynamics of first layer. Specifically we study cases where the coupling in system layer L1 is local with nonlocality in environment layer L2 and vice versa.

\begin{figure*}
\includegraphics[width=1.0\textwidth]{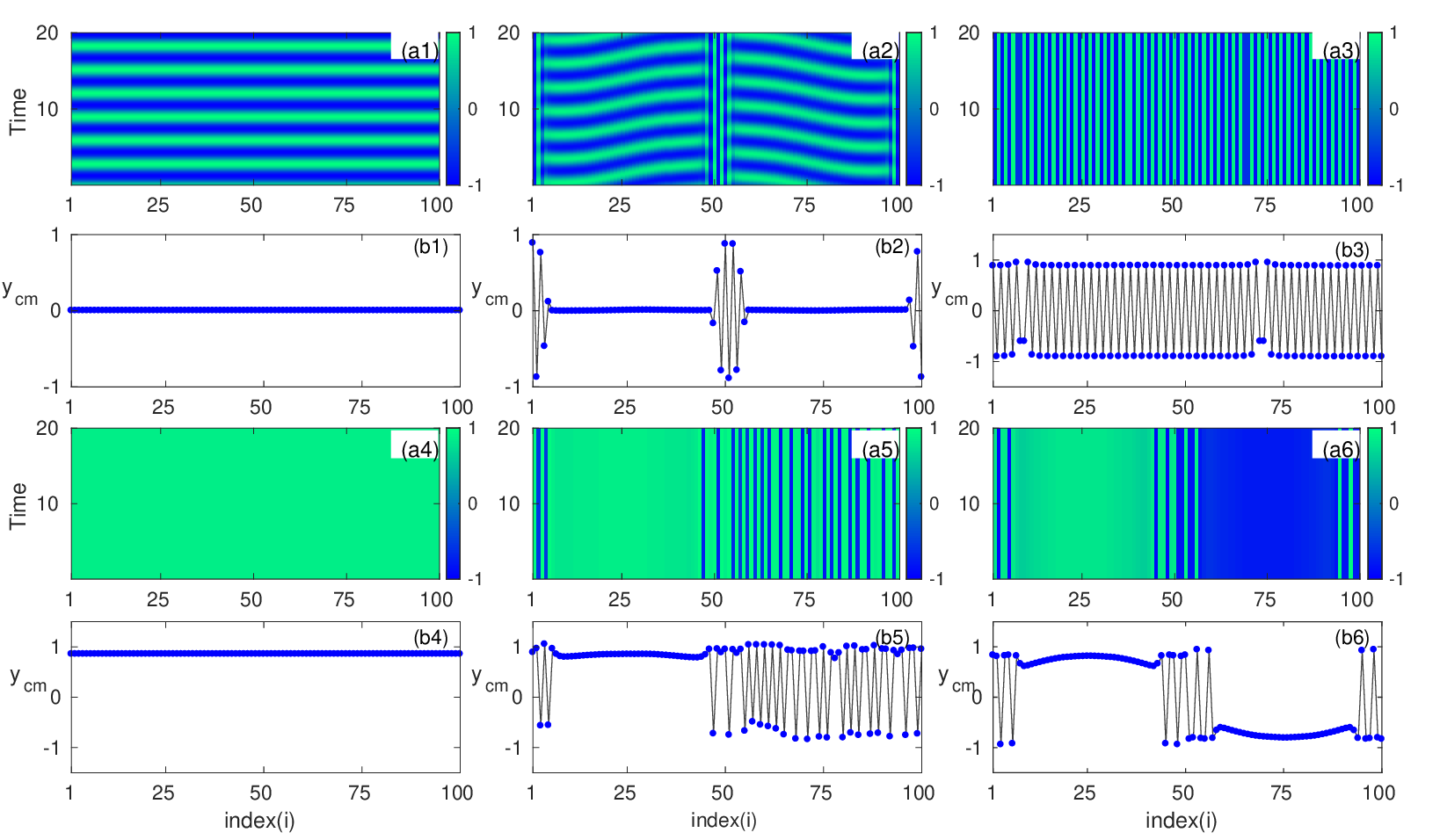}
\caption{Space-time dynamics for variable $y_i$ and corresponding center of mass averaged over one period of each oscillator at different value of coupling strength. (a1, b1) at $\epsilon=0.4$: synchronized
oscillation, (a2, b2) at $\epsilon=2$: amplitude chimera, (a3, b3) at $\epsilon=3$: inhomogeneous steady state, (a4, b4) at $\epsilon=4.5$: homogeneous steady state,(a5, b5) at $\epsilon=6.15$: one state chimera death(1-CD) and (a6, b6) at $\epsilon=6.5$: 2-state chimera death. The other parameters are set at $\omega=2$, $R_1=0.01$, $R_2=0.25$, $\sigma=10$, $\eta=10$, $\gamma=1$ and $N=100$.}
\label{fig2}
\end{figure*}

In the model of two layer network introduced above, the dynamic agents $s_i$ can be interpreted, based on context, in many different ways. They can be particle species that can freely diffuse in the surrounding medium and allow individual oscillators to communicate with each other. In the context of synthetic bacteria this dynamical agents $s_i$ can represent signalling molecules (called auto-inducers) which can freely diffuse in the local medium and in turn effect the collective dynamics of the cells~\cite{Garcia}. In the case of  Belousov-Zhabotinsky (BZ) reaction, $s_i$ represents the  chemical species that diffuse between autocatalytic beads~\cite{Taylor,Tinsley2010}. Similarly, for metabolic oscillations, $s_i$ represents the common metabolites that diffuse between cells~\cite{Schwab}.

In our study, we choose the initial conditions as follows. With random initial conditions and value of the parameters of the coupled SL oscillators chosen from the inhomogeneous steady-state (IHSS) regime, we find that the coupled system is divided into two domains, one located on the upper branch $(x_i,y_i)\approx(0.1, -0.85)$ and other on  the lower branch $(x_i,y_i)\approx(-0.1,0.85)$. So we distribute the initial states of SL oscillators around these two fixed points. The initial conditions of the first half of the oscillators are distributed around upper branch i.e. $(x_i, y_i)=(0.1+\Delta \xi, -0.85+\Delta \xi)$, where $i=1,2.....N/2$ and remaining half of the oscillators have initial conditions around lower branch i.e. $(x_i, y_i)=(-0.1+\Delta \xi, 0.85+\Delta \xi)$, where $i=\frac{N}{2}+1,.....N$. The initial conditions for $s_i$ are $0.01+ \Delta \xi$, where $i=1,2.....N$. The value of $\Delta$ is $0.1$, and $\xi$ is a function that gives uniformly distributed random numbers between $0$ and $1$ with zero mean. Throughout the study, the number of oscillators, N is taken as 100, and the dynamics of coupled oscillators, is studied by solving  Eq.~\ref{eq1}, using fourth--order Runge--Kutta method with a time step $0.01$. The first  $10^6$  values are discarded as transients in the study.

\subsection{Amplitude chimeras and chimera death: L1 with local and L2 with nonlocal interactions}

We first consider a case where all the SL oscillators on L1 are coupled to each other locally (i.e., $R_1=\frac{P_1}{N}= 0.01$) and dynamic agents on L2 are coupled to each other nonlocally with coupling radius $R_2=\frac{P_2}{N}=0.25$.  We study how the nonlocality or long range  interactions in L2 can induce and control chimera states in L1. We fix the value of $\sigma=10$, and  $\eta=10$, and vary the strength of interlayer coupling, $\epsilon$. 

In  Fig.~\ref{fig2}, we present the space-time plots for variable $y_i$, for the different values of $\epsilon$. For a value of $\epsilon=0.4$, the dynamics on system layer L1, shows synchronized oscillations, which is shown in Fig.~\ref{fig2} (a1).  By increasing the value of $\epsilon$  ($\epsilon=2$), we observe stable amplitude chimera, as shown in Fig.~\ref{fig2} (a2). This figure shows the existence of stable amplitude chimera plotted after discarding transients for a long time ($10^6$ time steps). We also calculate the center of mass for these two different values of $\epsilon$ using $y_{cm}=\int_{0}^{T} y_i(t)dt/T $, where $T=2\pi/\omega$ is the oscillation period for the $j_{th}$ oscillator. 
The center of mass values are plotted corresponding to  $\epsilon=0.4$ and $\epsilon=2$, in Fig.~\ref{fig2}(b1) and Fig.~\ref{fig2}(b2), respectively. From  Fig.~\ref{fig2}(b2), it is clear that when all the oscillators  are coherent in oscillations, $y_{cm}=0$, that is zero shift for the center of mass from the origin, while the systems oscillating with the incoherent region show shift in the values of center of the mass from the origin. 

When the interlayer coupling strength increased, we find the dynamics in L1, settles to different steady states.  Thus at $\epsilon=3$, L1 exhibits inhomogeneous steady-state (IHSS), as shown in  Fig.~\ref{fig2}(a3, c3), homogeneous steady-state (HSS) at $\epsilon=4.5$ (Fig.~\ref{fig2}(a4, b4)) etc. However for further increase of $\epsilon $, L1 stabilises to one state chimera death (1-CD) at $\epsilon=6.15$ and two-state chimera death (2-CD) at $\epsilon=6.5$, as shown in  Fig.~\ref{fig2}(a5, b5) and Fig.~\ref{fig2}(a6, b6) respectively.

\begin{figure}
\centering
\includegraphics[width=0.45\textwidth]{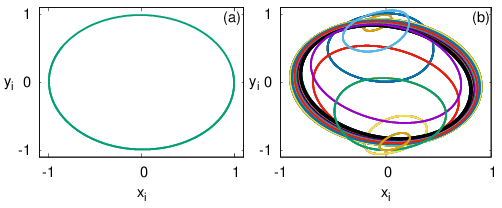}
\caption{Phase portraits of coupled SL oscillators (a) at $\epsilon=0.4$ showing synchronized oscillations and (b) at $\epsilon=2$ showing amplitude chimera. Here $\omega=2$, $R_1=0.01$, $R_2=0.25$, $\sigma=10$, $\eta=10$, $\gamma=1$, and $N=100$.}
\label{fig3}
\end{figure}

The phase portraits of coupled SL oscillators for the synchronized regime at $\epsilon=0.4$, and stable amplitude chimera regime at $\epsilon=2$ are plotted in Fig.~\ref{fig3}(a) and (b) respectively.

\subsection{Characterization of chimera states and their transitions}

\begin{figure}
\centering
\includegraphics[width=0.45\textwidth]{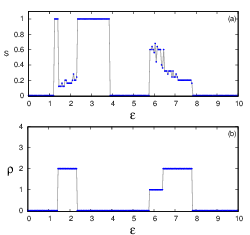}
\caption{(a) Strength of incoherence $S$ plotted against interlayer coupling strength $\epsilon$. $S=0$ indicates spatially synchronized state, $S=1$, completely desynchronized state and $0<S<1$ for chimera states (b) Discontinuity measure $\rho$ as a function of $\epsilon$. $\rho=0$ for coherent or incoherent state and unity indicates chimera state. $\rho$ is $1$, in figure indicates one-cluster chimera death, and $\rho=2$, two-states chimera death. Here $\omega=2$, $R_1=0.01$, $R_2=0.25$, $\sigma=10$, $\eta=10$, $\gamma=1$, and $N=100$.}
\label{fig4}
\end{figure}

In order to characterize the nature of chimera states, we calculate the strength of incoherence (S), as introduced by Gopal, et. al. ~\cite{Gopal}. This index will help us to distinguish chimera state from various other collective dynamical states such as the coherent state and incoherent state and can thus be used to study dynamical transitions in the system. We start by calculating  $w_{l,i} = x_{l,i}-x_{l,i+1}$,  where $l=1,2...d$ represents the dimension of individual units in the ensemble, $i =1,2,3, . . . ,N$. We divide the oscillators into $M$ bins of equal size $n = N/M$, and the local standard deviation $\sigma(m)$ is defined as

\begin{equation}
 \sigma_l(m)=\left\langle{\sqrt{\frac{1}{n}\sum_{j=n(m-1)+1}^{mn} [w_{l,j}-\bar{w}_{l,j}]^2}}\right\rangle_{t}, m=1,2...,M
\end{equation}

where $\bar{w}=\frac{1}{n}\sum_{j=n(m-1)+1}^{mn} w_{l,j}(t)$  and $\left\langle \cdots\right\rangle_t$ represents average over time. Now S is defined as, 

\begin{equation}
 S=1-\frac{\sum_{m=1}^{M}s_m}{M}, \quad s_m=\Theta (\delta-\sigma_l(m))
\end{equation}

where $\Theta(\cdot)$ is the Heaviside step function. $\delta$ is a  predefined threshold value, which is taken to be very small, usually fixed as a certain percentage of difference between $x_{l,i_{max}}$ and $x_{l,i_{min}}$. In the present study we take $M=20$ and $\delta=0.2$. In the incoherent domains $\sigma_l(m)$ has some finite value greater than $\delta$, hence the value of $s_m=0$, while in the coherent domains  $\sigma_l(m)$ is always zero, and hence $s_m= 1$. Consequently, $S$ takes the value $S=0$ for spatially synchronized state, $S=1$ for completely desynchronized state and take intermediate value between 0 and 1 (i.e. $0<S<1$) for chimera state or cluster state. The strength of incoherence (S) is shown in Fig.~\ref{fig4}(a) as a function of the interlayer coupling strength $\epsilon$ which indicates regions of spatial synchronization, non-synchronization and chimera states in L1.

\begin{figure}[t]
\centering
\includegraphics[width=0.43\textwidth]{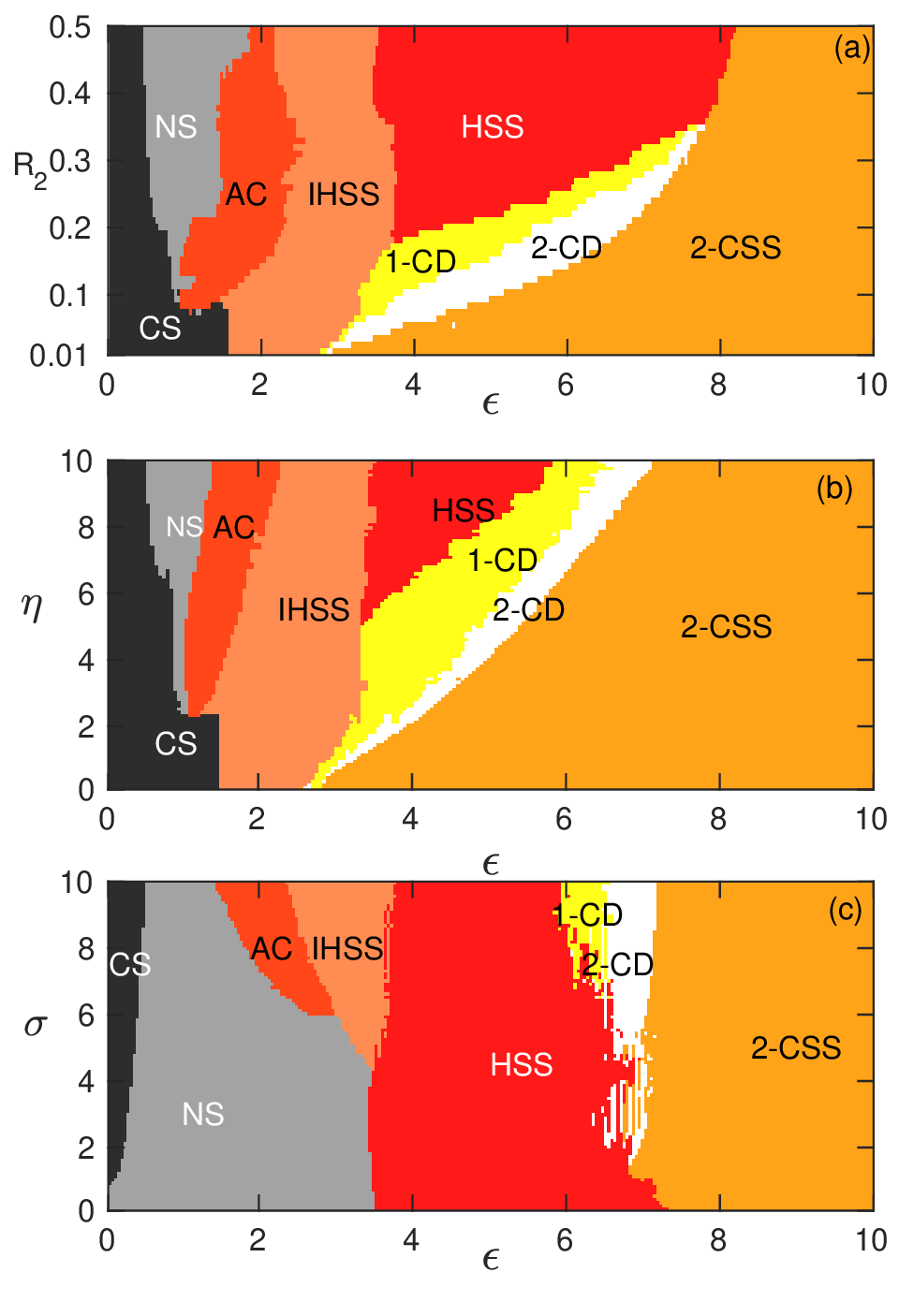}
\caption{Dynamical domains of $N$ coupled SL oscillators in the parameter plane (a) $\epsilon-R_2$ for $\eta=10$ and $\sigma=10$, (b) $\epsilon-\eta$ for $R_2=0.25$ and $\sigma=10$ and (c) $\sigma-\epsilon$ for $R_2=0.25$ and $\eta=10$.  Here CS, NS, AC, IHSS, HSS 1-CD, 2-CD, and 2-CSS represent complete synchronization, no-synchronization, amplitude chimera, inhomogeneous steady state, homogeneous steady state, one-state chimera death, two-state chimera death, and two clusters steady-state respectively. The other parameters are  $\omega=2$, $R_1=0.01$, $\gamma=1$, and $N=100$.}
\label{fig5}
\end{figure}

Also we characterize different types of multi-chimera states, using a discontinuity measure, which is based on the distribution of $s_m$. It is defined as~\cite{Premalatha},

\begin{equation}
 \rho=\frac{\sum_{i=1}^{M} |s_i-s_{i+1}|}{2}, \quad (s_{M+1}=s_1)
\end{equation}

The value of $\rho$ is zero for coherent or incoherent state and unity for chimera state. It takes positive integer value between $(1 < \rho \leq M/2)$ for multi-chimera states. Thus for one-cluster chimera death, the value of $\rho$ is $1$, and for two-states chimera death $\rho=2$ etc. The discontinuity measure $\rho$ is plotted as a function of $\epsilon$ in Fig.~\ref{fig4}(b). From this the region of coherent or incoherent states, one cluster chimera and two cluster chimera states can be identified clearly.

\begin{figure}[!ht]
\centering
\includegraphics[width=0.45\textwidth]{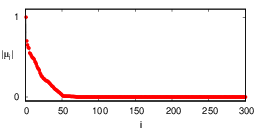}
\caption{ Floquet multipliers $|\mu_i|$ of $N=100$ coupled SL oscillators indicating stability of amplitude chimera state with $\epsilon=2.0$, $R_2=0.25$, $R_1=0.01$, $\gamma=1$,$\omega=2$, $\eta=10$, and $\sigma=10$. Here $i=1,2...3N$.}
\label{fig6}
\end{figure}

\begin{figure*}
\includegraphics[width=1.0\textwidth]{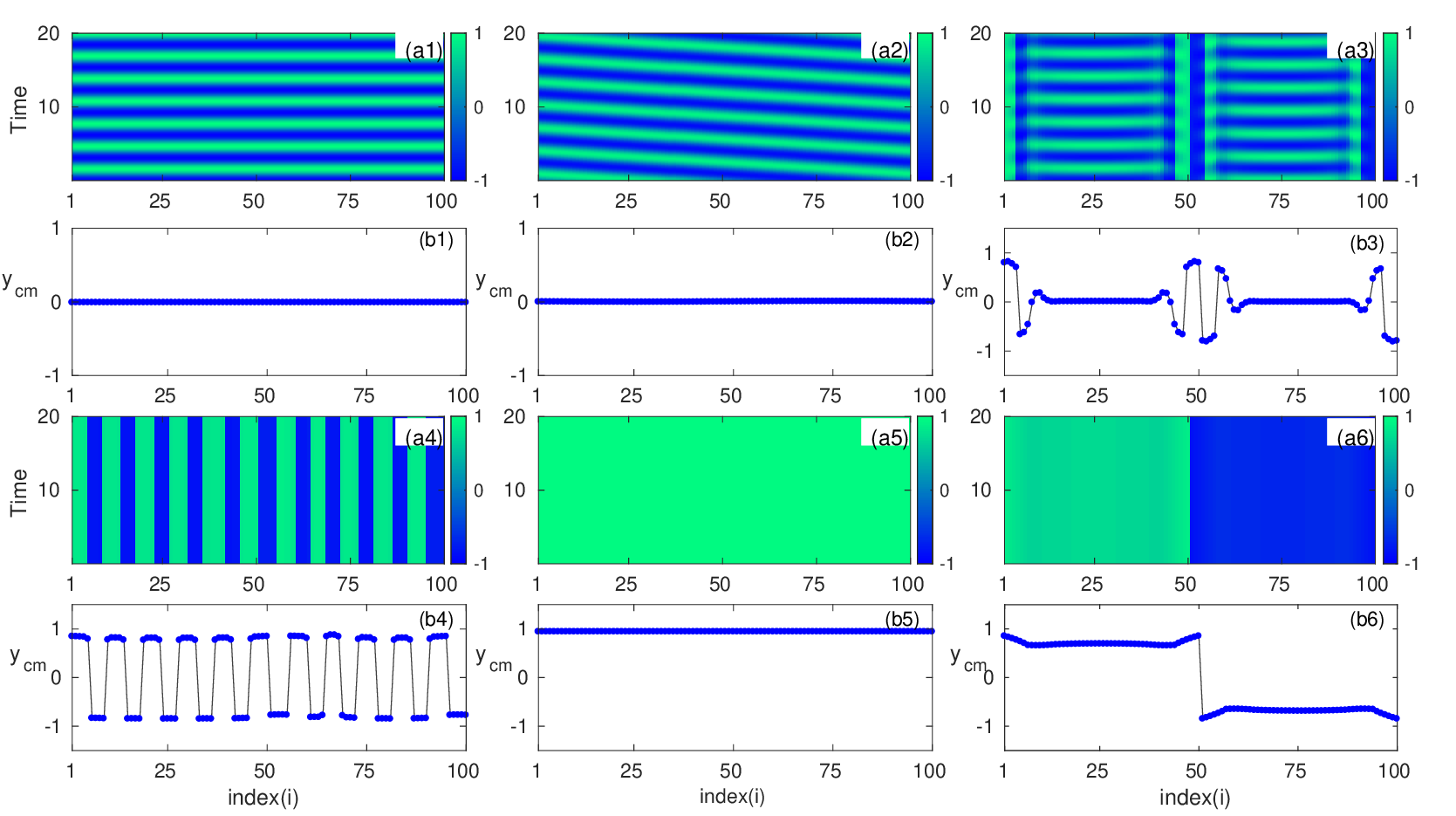}
\caption{Space-time plots for variable $y_i$ and corresponding center of mass averaged over one period of each oscillator at different value of coupling strength. (a1, b1) at $\epsilon=0.5$: oscillatory state and (a2, b2) at $\epsilon=1.3$: Traveling wave and (a3,c3) at $\epsilon=1.6$: Amplitude chimera. (a4, b4) at $\epsilon=3$: multi-cluster steady state, (a5, b5) at $\epsilon=6$: homogeneous steady state and (a6, b6) $\epsilon=8$: 2-cluster steady state. We set $\omega=2$, $R_1=0.05$, $R_2=0.36$, $\sigma=10$, $\eta=10$, $\gamma=1$, and $N=100$.}
\label{fig7}
\end{figure*} 

\subsection{Phase diagram on parameter planes}

We repeat the computation of the two measures, strength of incoherence, S and discontinuity measure $\rho$,  for a range of values of the strength and range of nonlocal coupling in L2 and present the various dynamical states possible on the two parameter phase diagram. We first fix $R_1=0.01$, $\eta=10$, $\sigma=10$ and plot the phase diagram $(\epsilon-R_2)$, which is shown in Fig.~\ref{fig5}(a).
In this figure, CS, NS, AC, IHSS, HSS 1-CD, 2-CD, and 2-CSS represent the complete synchronization, no-synchronization, amplitude chimera, inhomogeneous steady state, homogeneous steady state, one-state chimera death, two-state chimera death, and two clusters steady-state respectively. From this figure, we see that for non-local coupling radius in layer L2, $R_2<0.1$, the layer L1 shows synchronized oscillations. As the strength of coupling between layers $\epsilon$ increases, we see transitions to inhomogeneous steady-state (IHSS),  one state chimera death (1-CD), two-state chimera Death (2-CD), and for very strong coupling strength $\epsilon$ there is suppression of chimera giving two-cluster steady-state (2-CSS). In the two cluster steady state, dynamics on L1 is equally divided into two domains, one located on the upper branch, and the other is located at the lower branch.  For a larger range of coupling in L2, with $R_2>0.1$, L1 shows synchronized oscillations for weak coupling strength $\epsilon$. Increasing $\epsilon$, induces in L1 a series of interesting dynamics like amplitude chimera state (AC), inhomogeneous steady-state (IHSS), homogeneous steady-state (HSS, one state chimera death (1-CD), two-state chimera death and in the end, suppression of chimera to 2-CSS.

We study the possible emergent states on the parameter plane $(\epsilon-\eta)$ for the fixed values of the parameters $R_1=0.01$, $\sigma=10$, and $R_2=0.25$.  The corresponding phase diagram $(\epsilon-\eta)$ is shown in Fig.~\ref{fig5}(b). Here we observe that a stable amplitude chimera regime arises when $\eta>6$. Further, an increase of $\epsilon$ leads to an increase in the chimera death region. Thus nonlocal interactions in L2 induce chimera states in L1, but higher strength of coupling or increase in range of nonlocality can suppress chimera. We also plot the parameter plane $(\sigma-\epsilon)$ for the fixed values of the parameters $R_1=0.01$, $\eta=10$, and $R_2=0.25$ in Fig.~\ref{fig5}(c). Here we observe stable amplitude chimera only for higer value of $\sigma$. We can also see that 1-CD state arises when $\sigma>8$.

\subsection{Stability of the amplitude chimera states}
 We apply the Floquet theory~\cite{Sathiyadevi2, Tumash} to check the stability of amplitude chimera state. For this,we derive equations for perturbations from the chimera state starting from Eqn~\ref{eq1} as: 

\begin{eqnarray}
\dot{\xi}_{i}&=&a_1\xi_{i}-(\omega+ 2x_i^*y_i^*)\lambda_i+\frac{\sigma}{2P_1}\sum_{j=i-P_1}^{i+P_1}(\xi_{j}-\xi_{i})+\epsilon \kappa_i \nonumber\\
\dot{\lambda}_i&=&a_2\lambda_i+(\omega- 2x_i^*y_i^*)\xi_i \nonumber\\
\dot{\kappa}_i &=& -\gamma \kappa_i-\epsilon \xi_{i}+\frac{\eta}{2P_2}\sum_{j=i-P_2}^{i+P_2}(\kappa_j-\kappa_i)
\label{eq5}
\end{eqnarray}

where $a_1=(1-3x_i^{*2}-y_i^{*2})$  and $a_2=(1-x_i^{*2}-3y_i^{*2})$.  $x_i^*$, $y_i^*$ and $s_i^*$ are the solutions of the amplitude chimera  and $\xi_i$, $\lambda_i$ and $\kappa_i$ are the perturbations. Integrating the above equation for one time period $T=2\pi/\omega$, we can construct the monodromy matrix. Then we calculate the eigenvalues of the monodromy matrix, to get the Floquet multipliers ($\mu_i$)~\cite{Sathiyadevi2,Tumash}, that characterize the stability of a periodic orbit. If all $|\mu_i|$ are less then one (except for the Goldstone mode i.e. $|\mu_1|=1$) the corresponding periodic orbit is stable. In Fig~\ref{fig6} we  plot the values of all Floquet multipliers for  amplitude chimera state.  Since all values of $|\mu_i|$ is less than one except $|\mu_1|=1$, it is clear that the periodic orbits constituting the amplitude chimera are stable.

\begin{figure}
\centering
\includegraphics[width=0.45\textwidth]{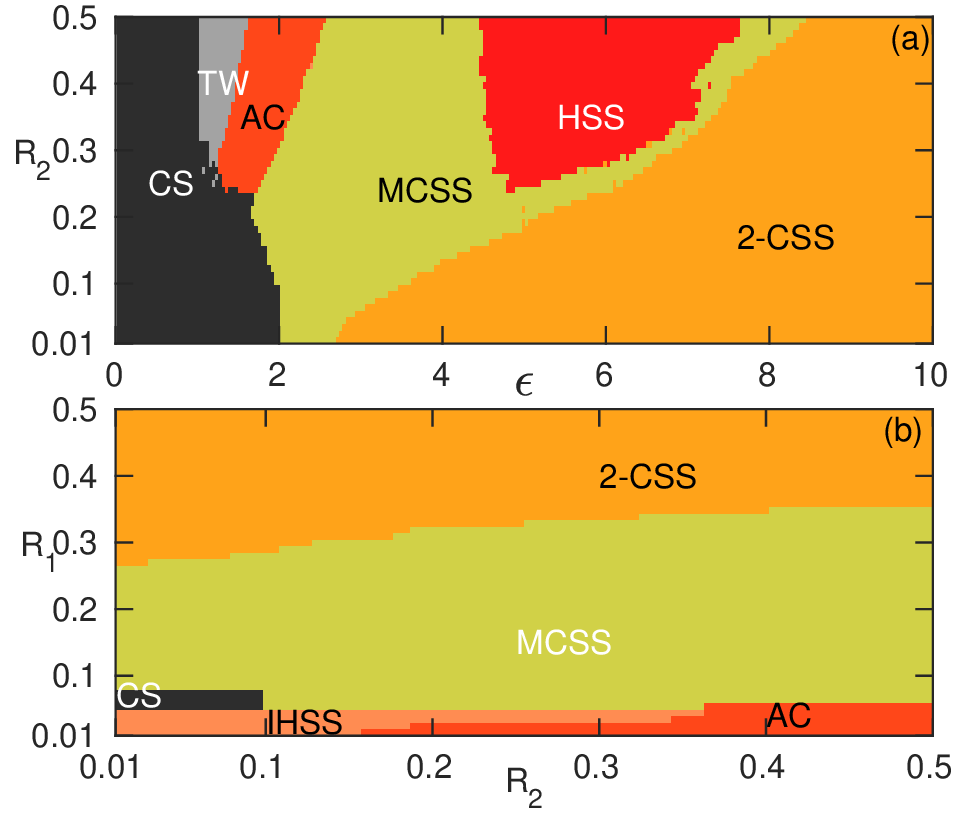}
\caption{Dynamical states of coupled SL oscillators in the parameter plane (a) $(\epsilon-R_2)$ for $R_1=0.05$ and (b) $(R_2-R_1)$ for $\epsilon=2$ with $\omega=2$, $\sigma=10$, $\eta=10$, $\gamma=1$, and $N=100$. Here CS, TW, AC, MCSS, IHSS, HSS  and 2-CSS represent the complete synchronization, traveling wave, amplitude chimera, multi-cluster steady state, inhomogeneous steady state, homogeneous steady state, and two clusters steady-state respectively.  }
\label{fig8}
\end{figure}

\subsection{Suppression of chimera: L1 and L2 with nonlocal interactions }

We now consider the case when $R_1>0.01$ i.e. the ensemble of SL oscillators interact directly through nonlocal coupling in layer L1 while the dynamic agents also interact nonlocally in the layer 2. We first fix $R_1=0.05$, $R_2=0.36$, $\sigma=10$ and $\eta=10$, and plot space-time plot for different values of $\epsilon$ (Fig.~\ref{fig7}). For weak interlayer coupling strength at $\epsilon=0.5$ L1 exhibits complete synchronized oscillations, as shown in Fig.~\ref{fig7}(a1, b1). For a higher value of $\epsilon=1.3$, we observe traveling wave(TW)  dynamics in L1 (Fig.~\ref{fig7}(a2,b2)). However further increase to $\epsilon=1.6$, results in stable amplitude chimera as is clear from Fig.~\ref{fig7}(a3,b3) and at $\epsilon=3$ we see multi-cluster steady state (MCSS)(Fig.~\ref{fig7}(a4,b4)). We observe HSS and two clusters steady state (2-CSS) for higher values of $\epsilon=6$ and $\epsilon=8$, which are shown in Fig.~\ref{fig7}(a5,b5) and Fig.~\ref{fig7}(a6,b6) respectively.

For this coupling scenario, we plot phase diagram $(\epsilon-R_2)$ keeping other parameter values as $\omega=2$, $\sigma=10$ and $\eta=10$. In Fig.~\ref{fig8}(a) the phase diagram in the parameter space $(\epsilon-R_2)$ are shown for $R_1=0.05$. It shows the regions of  traveling wave (TW), stable amplitude chimera(AC) and HSS states that arise for higher value of coupling radius $R_2$. In this case we also see multi-cluster steady state (MCSS) and two cluster steady state (2-CSS) with increase of $R_2$. We also present a phase diagram in the parameter space $(R_2-R_1)$ for $\epsilon=2$ in Fig.~\ref{fig8}(b). Here we observe IHSS and AC state at very small value of $R_1$ . In the parameter space we also have MCSS when $R_1 > 0.04$. We also observe two clusters steady state (2-CSS) for higher value of $R_1$ . From this parameter space it is clear that amplitude chimera state occurs only for small value of $R_1$ .

\begin{figure}
\centering
\includegraphics[width=0.45\textwidth]{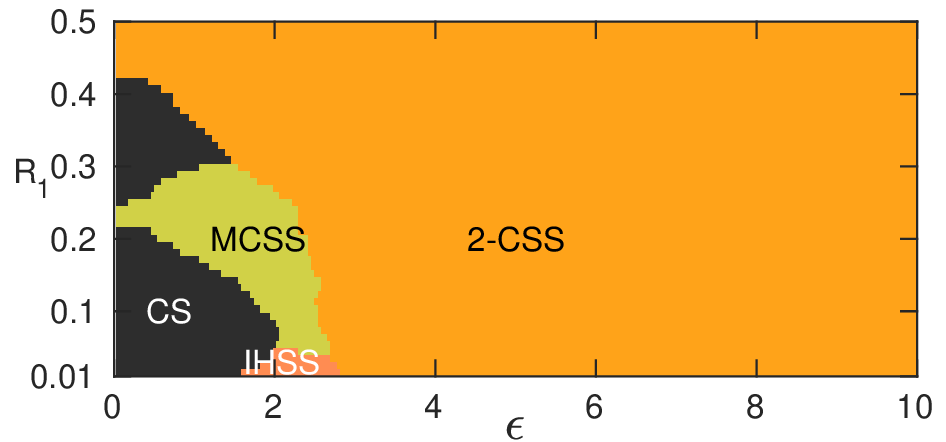}
\caption{Dynamical states of coupled SL oscillators in the parameter plane (a) ($\epsilon-R_1$) for $R_2=0.01$ with $\omega=2$, $\sigma=10$, $\eta=10$, $\gamma=1$, and $N=100$. Here CS, IHSS, MCSS, and 2-CSS represent complete synchronization, inhomogeneous steady state, multi-cluster steady state, and 2-cluster steady state respectively. }
\label{fig9}
\end{figure}

\subsection{L1 with nonlocal and L2 with local interactions }

We also consider the case where oscillators in L1 are coupled to each other nonlocally (i.e., $R_1>0.01$) while dynamical agents in L2 are coupled locally (i.e., $R_2=0.01$). The corresponding phase diagram $(\epsilon-R_1)$, is shown in Fig.~\ref{fig9}. In this case, we do not see chimera states even though L1 has nonlocal couplings.  When the value of $R_1$ is small L1 shows a transition from complete synchronized state to IHSS state as the coupling strength increases, and further transition from IHSS to 2-CSS. For an increase in the range of coupling, L1 mostly shows only 2-CSS. 

\section{Conclusion}
In summary, we present emergent behavior in a two-layer network, in which layer L1 is formed by an ensemble of identical oscillators interacting through a local coupling, and layer L2 forms another network of dynamic agents with nonlocal coupling among them. The two layers are put in a feedback loop so that they can mutually influence their dynamics. For the specific case of coupled Stuart-Landau oscillators with the limit cycle dynamics, we show how the layer L2 functioning as a dynamic environment can be tuned to control the dynamics in L1. 

Our study indicates that the long-range interactions in L2, can induce stable amplitude chimera and chimera death in L1, even when L1 has only short-range or local interactions. With an increase in coupling strength between the layers and range of interaction in L2, different types of steady-states such as homogeneous steady state, inhomogeneous steady-state, and two- clusters- steady-state are found to occur. In the chimera death regime, we find one state chimera death and two-state chimera death. We compute two quantifiers, strength of incoherence to identify occurrence of chimera and discontinuity measure to distinguish different types of chimeras and chimera death states. We use them to identify regions of different emergent dynamics in phase diagrams on parameter planes. In all types of emergent behaviour, the dynamics in the layer L2 matches that of layer L1, and both layers exhibit spatially coherence and the temporarily phase-shifted dynamics. On repeating the study for larger $N$ values, we observe qualitatively similar results.

We also present two other possibilities, where both layers have nonlocal interactions as well as the case where oscillators in L1 interact through nonlocal coupling and are coupled to L2 that has only local interactions. In the former case, we observe traveling waves and stable amplitude chimera for weak coupling strength but mostly multi-clusters steady-state (MCSS) and HSS states. In the latter case, the system settles to 2-cluster steady-state and multi-cluster steady states, even though the layer L1 has nonlocal interactions. 

In the limiting case of no coupling in layer L1 but nonlocal coupling in L2, we see homogeneous steady-state and 2-cluster steady states. Similarly, with L1 having nonlocal coupling but L2 has no coupling, only multi-cluster steady states and 2- cluster steady states are seen to occur.  In both cases, synchronized states occur for low coupling strengths.

The model of interacting two layer networks presented is very generic and can be applied to a wide class of systems ranging from chemical oscillators~\cite{Taylor,Tinsley2010}, synthetic genetic~\cite{Ullner} and neuronal systems~\cite{Dotson}, systems of bacteria communicate with each other through chemical species~\cite{Ullner}. In general, the study illustrates how dynamics in one layer can be controlled by tuning that in the other, even when both have different intrinsic dynamics and different ranges of interactions.


\end{document}